\documentstyle[prd,aps,12pt,epsf]{revtex}
\begin{document}  
\baselineskip=2pc  
\draft
\title{Vacuum polarization of scalar fields near Reissner-Nordstr\"{o}m black holes and the resonance behavior in field-mass dependence}
\author{Akira Tomimatsu\thanks{Email: atomi@allegro.phys.nagoya-u.ac.jp} and Hiroko Koyama\thanks{Email: hiroko@allegro.phys.nagoya-u.ac.jp}}
\address{Department of Physics, Nagoya University, Nagoya 464-8602, Japan}
\date{31 January, 2000}
\maketitle
\begin{abstract}
We study vacuum polarization of quantized massive scalar fields $\phi$ in equilibrium at black-hole temperature in Reissner-Nordstr\"{o}m background. By means of the Euclidean space Green's function we analytically derive the renormalized expression $<\phi^{2}>_{H}$ at the event horizon with the area $4\pi r_{+}^{2}$. It is confirmed that the polarization amplitude $<\phi^{2}>_{H}$ is free from any divergence due to the infinite red-shift effect. Our main purpose is to clarify the dependence of $<\phi^{2}>_{H}$ on field mass $m$ in relation to the excitation mechanism. It is shown for small-mass fields with $mr_{+}\ll1$ how the excitation of $<\phi^{2}>_{H}$ caused by finite black-hole temperature is suppressed as $m$ increases, and it is verified for very massive fields with $mr_{+}\gg1$ that $<\phi^{2}>_{H}$ decreases in proportion to $m^{-2}$ with the amplitude equal to the DeWitt-Schwinger approximation. In particular, we find a resonance behavior with a peak amplitude at $mr_{+}\simeq 0.38$ in the field-mass dependence of vacuum polarization around nearly extreme (low-temperature) black holes. The difference between Scwarzschild and nearly extreme black holes is discussed in terms of the mass spectrum of quantum fields dominant near the event horizon.  
\end{abstract}

\section{introduction}
\label{intro}
Quantum behaviors of matter fields in black hole spacetimes have been extensively studied for understanding the various physical effects. In particular, the existence of a state of quantum fields in equilibrium at a finite temperature near the event horizon has attracted much attention, because it clearly represents the thermodynamic properties of stationary black holes. The problem of vacuum polarization for this Hartle-Hawking state \cite{HH} may be described in terms of the Euclidean space Green's function $G_{E}(x,x')$, which is periodic with respect to the Euclidean time $\tau=it$. If one considers a quantized scalar field $\phi$, the vacuum polarization $<\phi^{2}(x)>$ can be calculated by using the equality
\begin{equation}
<\phi^{2}(x)> \ = \ \rm{Re}\{\lim_{x'\rightarrow x} G_{E}(x,x')\} \ , \label{GE}
\end{equation}
in which the renormalised expression is derived through the method of point splitting.

It is well-known that the black-hole temperature $T$ defined as the inverse of the period of $G_{E}(x,x')$ is proportional to the surface gravity $\kappa$ on the event horizon as follows,
\begin{equation}
T \ = \ \kappa/2\pi \ . 
\end{equation}
(Throughout this paper we use units such that $G=c=\hbar=k_{B}=1$.) If the origin of the vacuum polarization $<\phi^{2}(x)>$ is claimed to be purely induced by the finite black-hole temperature, the amplitude should decrease toward zero in the extreme black-hole limit $\kappa\rightarrow 0$. In fact, we can see this behavior of $<\phi^{2}>$ by applying the analytical approximation of the renormalized value obtained by Anderson, Hiscock and Samuel \cite{An95} to Reissner-Nordstr\"{o}m background, for which the analytic continuation of the metric into Euclidean space is given by 
\begin{equation}
ds^{2} \ = \ f(r)d\tau^{2}+f^{-1}(r)dr^{2}+r^{2}d\theta^{2}+r^{2}\sin^{2}\theta d\varphi^{2} \ , \label{RN}
\end{equation}
where $f=(r-r_{+})(r-r_{-})/r^{2}$, and using mass $M$ and charge $Q$ parameters of the black hole, we have 
\begin{equation}
r_{\pm} \ = \ M\pm\sqrt{M^{2}-Q^{2}} \ . 
\end{equation}
For massless scalar fields the analytical approximation denoted by $<\phi^{2}>_{T}$ reduces to 
\begin{equation}
<\phi^{2}(r)>_{T} \ = \ \frac{\kappa^{2}}{48\pi^{2}}\times\frac{(r+r_{+})(r^{2}+r_{+}^{2})}{r^{2}(r-r_{-})} \ . \label{massless}  
\end{equation}
Therefore, in nearly extreme Reissner-Nordstr\"{o}m spacetime such that $\kappa r_{+}=(r_{+}-r_{-})/(2r_{+})\ll 1$, the vacuum polarization of massless fields is strongly suppressed. (This is also justified by the result of Frolov \cite{Fr82} estimated at the event horizon $r=r_{+}$.) 

Such a excitation of vacuum polarization induced by finite black-hole temperature is an important aspect of quantum matter fields in black hole backgrounds, and it may remain valid for massive scalar fields too. Then, field mass $m$ will just play a role of suppressing the amplitude of $<\phi^{2}>$ in comparison with massless fields. In this paper, however, we would like to emphasize another remarkable effect due to field mass, which we call mass-induced excitation as a remaining part of $<\phi^{2}>$ in the low-temperature limit $T\rightarrow 0$. Note that massive fields can have a characteristic correlation scale corresponding to the Compton wavelength $1/m$. Our purpose is to show that nearly extreme (low-temperature) black holes can enhance the excitation of quantum fields with the Compton wavelength $1/m$ of order of the black hole radius (i.e., $mr_{+}\sim 1$). This mass-induced excitation may be expected as a result of wave modes in resonance with the potential barrier surrounding a black hole, for which the tail part of $<\phi^{2}>$ in the large-mass limit $mr_{+}\gg1$ is generated with the amplitude decreasing in proportion to $1/m^{2}$ \cite{Fr84,An90} according to the DeWitt-Schwinger approximation developed by Christensen \cite{Ch}.

In this paper our investigation is focused on Reissner-Nordstr\"{o}m background as the simplest example which allows us to consider the low-temperature limit $\kappa r_{+}\ll 1$ keeping an arbitrary value of $mr_{+}$. (The black hole temperature and the field mass are measured in unit of the inverse of a fixed black hole radius $r_{+}$. In Schwarzschild background with $\kappa r_{+}=1/2$ we cannot discuss the field-mass dependence of $<\phi^{2}>$ in such a low-temperature limit, and any resonance behavior of the polarization amplitude $<\phi^{2}>$ at $mr_{+}\sim 1$ will become obscure by virtue of a contamination of the temperature-induced excitation in the mass range of $mr_{+}\ll1$ \cite{An89}.) Then, following the analysis given by Anderson and his collaborators \cite{An95,An90}, we compute the vacuum polarization of massive scalar fields, for which we have the analytical approximation of the form
\begin{equation}
<\phi^{2}>_{ap} \ = \ <\phi^{2}>_{T}+<\phi^{2}>_{m^{2}} \ , \label{AP}
\end{equation}
Here the additional contribution from field mass becomes
\begin{equation}
<\phi^{2}>_{m^{2}} \ = \ \frac{m^{2}}{16\pi^{2}}\{1-2\gamma-\ln(\frac{m^{2}f}{4\kappa^{2}})\} \ , \label{massive}
\end{equation}    
with Euler's constant $\gamma$. Unfortunately, this field-mass term contains a logarithmic divergence at the event horizon $r=r_{+}$. Therefore, in Sec. II we develop the technique of analytical calculation to cancel such a divergent term, by paying the price that $<\phi^{2}>$ is evaluated only near the event horizon. It is checked in Sec. III that the renormalized value of $<\phi^{2}>$ at the event horizon becomes identical, up to the leading terms of order of $1/m^{2}r_{+}^{2}$, with the result derived by DeWitt-Schwinger expansion in the large-mass limit. In Sec. IV, using the small-mass approximation $mr_{+}\ll 1$, we show the tendency of temperature-induced excitation to be suppressed with incerase of field mass. We find in Sec. V the mass-induced enhancement of the polarization amplitude $<\phi^{2}>$, by giving explicitly the dependence on field mass in the low-temperature limit $\kappa r_{+}\ll 1$. The final section summarizes the results representing a remarkable difference of field-mass dependence of the polarization amplitude for scalar fields in equilibrium at various black-hole temperatures.

\section{correction to the wkb approximation}
\label{wkb}
Let us start from a brief introduction of the method to compute the renormalized value of $<\phi^{2}>$ in Reissner-Nordstr\"{o}m background (\ref{RN}), which has been developed by Anderson and his collaborators \cite{An95,An90}. Using Eq.\ (\ref{GE}) for a massive scalar field $\phi$ obeying the equation
\begin{equation}
(\Box-m^{2})\phi(x) \ = \ 0 \ , \label{wave}
\end{equation}
the unrenormalized expression is given by
\begin{equation}
<\phi^{2}(r)> \ = \ \lim_{\epsilon\rightarrow0}\{\frac{\kappa}{4\pi^{2}}\sum_{n=0}^{\infty}c_{n}\cos(n\kappa\epsilon)A_{n}(r)\} \ , \label{unren}
\end{equation}  
where $c_{0}=1/2$ and $c_{n}=1$ for $n\geq1$. The separation of two points in $G_{E}(x,x')$ is chosen to be only in time as $\epsilon\equiv\tau-\tau'$, and the radial part $A_{n}(r)$ for each quantum number $n$ is given by the sum of radial modes $p_{nl}(r)$ and $q_{nl}(r)$,  
\begin{equation}
A_{n}(r) \ = \ \sum_{l=0}^{\infty}\{(2l+1)p_{nl}(r)q_{nl}(r)-\frac{1}{r\sqrt{f}}\} \ , \label{sum}
\end{equation}
where $l$ is the angular-momentum quantum number, and the subtraction term $1/r\sqrt{f}$ is necessary for removing the divergence in the sum over $l$. The radial mode $q_{nl}$ satisfies the equation
\begin{equation}
\frac{d^{2}q_{nl}}{dr^{2}}+\frac{1}{r^{2}f}\frac{d(r^{2}f)}{dr}\frac{dq_{nl}}{dr}-\{\frac{n^{2}\kappa^{2}}{f^{2}}+\frac{l(l+1)+m^{2}r^{2}}{fr^{2}}\}q_{nl} \ = \ 0 \ , \label{qnl}
\end{equation}
and it is chosen to be regular at $r=\infty$ and divergent at $r=r_{+}$. The same equation is satisfied by $p_{nl}$, which is chosen to be well-behaved at $r=r_{+}$ and divergent at $r=\infty$. 

The WKB approximation for the modes may be used to calculate the mode sums (\ref{sum}), by assuming the forms 
\begin{equation}
p_{nl} \ = \ \frac{1}{(2r^{2}W)^{1/2}}\exp(\int (W/f)dr) \ , \label{WKBp}
\end{equation}
and
\begin{equation}
q_{nl} \ = \ \frac{1}{(2r^{2}W)^{1/2}}\exp(-\int (W/f)dr) \ , \label{WKBq}
\end{equation}
where the zeroth-order solution is chosen to be
\begin{equation}
W^{2} \ \simeq \ n^{2}\kappa^{2}+\{(l+\frac{1}{2})^{2}+m^{2}r^{2}\}\frac{f}{r^{2}} \ . \label{zeroth}
\end{equation}
To renormalize $<\phi^{2}>$ in the limit $\epsilon\rightarrow 0$ of point splitting, we subtract the counterterms $<\phi^{2}>_{DS}$ generated from the DeWitt-Schwinger expansion of $<\phi^{2}>$,
\begin{equation}
<\phi^{2}>_{DS} \ = \ \frac{1}{8\pi^{2}\sigma}+\frac{m^{2}}{16\pi^{2}}\{-1+2\gamma+\ln(\frac{m^{2}|\sigma|}{2})\}+\frac{1}{96\pi^{2}}R_{ab}\frac{\sigma^{a}\sigma^{b}}{\sigma} \ , \label{DS}
\end{equation}
where $\sigma$ is equal to one half the square of the distance between the two points $x$ and $x'$, and $\sigma^{a}\equiv\nabla^{a}\sigma$. Then, for the renormalized value defined by
\begin{equation}
<\phi^{2}>_{ren} \ = \ <\phi^{2}>-<\phi^{2}>_{DS} \ , \label{ren}
\end{equation}
we can arrive at the analytical approximation (\ref{AP}), if the second-order WKB approximation for $W$ is used in the mode sums for $n\geq1$ \cite{An95,An90}.
 
Though Eq.\ (\ref{AP}) can clearly show a spatial distribution of the vacuum polarization, the validity is rather restrictive. For example, in the asymptotically flat region $r\rightarrow\infty$ it fails to give the expected dependence on field mass. It is instructive for later discussions to calculate precisely $<\phi^{2}>_{ren}$ of thermal fields in equilibrium at a temperature $T$ in flat background (corresponding to $f=1$), following the method of the Euclidean space Green's function $G_{E}(x,x')$. Denoting $T$ by $\kappa/2\pi$, we obtain the exact solutions for $p_{nl}$ and $g_{nl}$ in flat background as follows,
\begin{equation}
p_{nl} \ = \ \frac{1}{r^{1/2}}I_{l+\frac{1}{2}}(r\sqrt{m^{2}+n^{2}\kappa^{2}}) \ , 
\end{equation}
and
\begin{equation}
q_{nl} \ = \ \frac{1}{r^{1/2}}K_{l+\frac{1}{2}}(r\sqrt{m^{2}+n^{2}\kappa^{2}}) \ , 
\end{equation}
and the mode sum over $l$ in $A_{n}$ results in  
\begin{equation}
A_{n} \ = \ -\sqrt{m^{2}+n^{2}\kappa^{2}} \ . 
\end{equation}
If we use the Plana sum formula for a function $g(k)$
\begin{equation}
\sum_{j=k}^{\infty}g(j) \ = \ \frac{1}{2}g(k)+\int_{k}^{\infty}g(x)dx+i\int_{0}^{\infty}
\frac{dx}{e^{2\pi x}-1}[g(k+ix)-g(k-ix)] \ , \label{Plana}
\end{equation}
the unrenormalized value is written by the integral form 
\begin{equation}
<\phi^{2}> \ = \ \lim_{\epsilon\rightarrow0}\{\frac{\kappa}{4\pi^{2}}[-\int_{0}^{\infty}dn\cos(n\kappa\epsilon)\sqrt{m^{2}+n^{2}\kappa^{2}}+\int_{m/\kappa}^{\infty}\frac{2dn}{e^{2\pi n}-1}\sqrt{\kappa^{2}n^{2}-m^{2}}]\} \ . \label{flat}
\end{equation}  
The first term in the right-hand side of Eq.\ (\ref{flat}) is completely canceled by the subtraction of the DeWitt-Schwinger counterterms (\ref{DS}), in which we have $\sigma=-\epsilon^{2}/2$, while the second term gives the renormalized value $<\phi^{2}>_{ren}$ in flat background, which for massless fields reduces to
\begin{equation}
<\phi^{2}>_{ren} \ = \ T^{2}/12 \ , 
\end{equation}
and becomes equal to Eq.\ (\ref{AP}) estimated in the asymptotically flat region. However, in the large-mass limit $m\gg\kappa$, we obtain
\begin{equation}
<\phi^{2}>_{ren} \ = \ m^{1/2}(T/2\pi)^{3/2}e^{-m/T} \ , 
\end{equation}
because the second integral over $n$ in Eq.\ (\ref{flat}) should run from the large lower limit $m/\kappa\gg 1$ to infinity. This leads to a crucial difference from the approximated form (\ref{AP}), for which $A_{n}$ is expressed in inverse powers of $n\kappa$ such that
\begin{equation}
A_{n} \ \simeq \ -\frac{n\kappa}{f}+(\frac{1}{12r^{2}}-m^{2})/2n\kappa \ , \label{expansion}
\end{equation}
as a result of the mode sum over $l$ using the zeroth-order solution (\ref{zeroth}) for $W$. It is clear that the sum of such an expansion form of $A_{n}$ over $n\geq1$ misses the exponential behavior $e^{-2\pi m/\kappa}$ of $<\phi^{2}>_{ren}$ in the asymptotically flat region. 

Now let us turn our attention to vacuum polarization at the event horizon $f=0$, which is the main concern in this paper. Fortunately, we can claim that the above-mentioned deviation of Eq.\ (\ref{AP}) from the precise estimation becomes irrelevant, if we consider the limit $f\rightarrow 0$. This is because owing to the redshift factor $f$ in $W$ the expansion (\ref{expansion}) remains valid even for a large mass $m\geq\kappa$, by keeping the condition $m\sqrt{f}/\kappa\ll1$. Then, concerning vacuum polarization of massive fields at the event horizon, we can use Eq.\ (\ref{AP}) to show the dependence of $<\phi^{2}>_{ren}$ on $m$. Of course, one may point out another crucial problem that Eq.\ (\ref{AP}) contains a logarithmic divergence at $r=r_{+}$. However, this singular behavior is due to the sum of $A_{n}$ over the limited range of $n\geq1$. Because the expansion form (\ref{expansion}) also breaks down for $n=0$, the contribution of $A_{0}$ to $<\phi^{2}>_{ren}$ is omitted in the calculation of Eq.\ (\ref{AP}). We would like to clarify an important role of the $n=0$ mode to give a regular value at the event horizon for the renormalized vacuum polarization $<\phi^{2}>_{ren}$ (and also for the renormalized stress-energy tensor $<T_{ab}>_{ren}$). 

To this end we propose the procedure to treat more precisely the mode sum over $l$ in $A_{n}$ at the event horizon, which is applicable to the lower $n$ modes. Note that near the event horizon the exact solution for $q_{nl}$ should have the expansion form  
\begin{equation}
q_{nl} \ = \ z^{n/2}\ln z\sum_{s=0}^{\infty}\alpha_{s}z^{s}+z^{-n/2}\sum_{s=0}^{\infty}\beta_{s}z^{s} \ , \label{smallz} 
\end{equation}
with some coefficients $\alpha_{s}$ and $\beta_{s}$. The rescaled radial coordinate $z$ is defined by $z\equiv(r-r_{+})/r_{+}\ll 1$. This expansion form is not useful to calculate $A_{n}$ at the event horizon, because the sums over $l$ should be done without expanding in powers of $z$ for requiring the convergence. Then, the important points to be mentioned here are the existence of the logarithmic term $z^{n/2}\ln z$ and the power-law behavior $z^{-n/2}$ dominant for $n\geq1$ in the limit $z\rightarrow 0$ (except for the $n=0$ mode in which the logarithmic term becomes dominant). For the modes $p_{nl}$ regular at the event horizon the dominant power-law behavior is given by $z^{n/2}$, and the WKB forms (\ref{WKBq}) and (\ref{WKBp}) for $q_{nl}$ and $p_{nl}$ remain exact up to these dominant power-law terms. Hence, the value of $A_{n}$ for $n\geq 1$ is exactly given by the WKB calculation in the limit $z\rightarrow0$, and we will obtain a precise value of $<\phi^{2}>_{ren}$ at the event horizon by taking account of the additional correction $A_{0}$ to Eq.\ (\ref{AP}).

To resolve the problem of logarithmic divergence, however, it is important to note that the WKB form for $q_{nl}$ fails to give the logarithmic behavior, which should play the role of canceling the logarithmic term contained in the DeWitt-Schwinger renormalization counterterms. (Because the leading logarithmic behavior in $A_{n}$ would be $z^{n}\ln z$, the value of $<\phi^{2}>_{ren}$ can become regular at the event horizon only by considering a more precise treatment of the $n=0$ mode beyond the WKB level, while the same analysis for the $n=1$ mode is also necessary to obtain a regular value of $<T_{a}^{b}>_{ren}$.) Hence, our key approach is to study the modified Bessel forms for the modes instead of the WKB forms as follows, 
\begin{equation}
p_{nl} \ = \ (\frac{\chi}{r^{2}w})^{1/2}I_{n}(\chi) \ , 
\end{equation}
and
\begin{equation}
q_{nl} \ = \ (\frac{\chi}{r^{2}w})^{1/2}K_{n}(\chi) \ , 
\end{equation}
where we have
\begin{equation}
\chi \ = \ \int_{r_{+}}^{r}(w/f)dr \ , 
\end{equation}
for which it is easy to check the validity of the Wronskian condition
\begin{equation}
p_{nl}\frac{dq_{nl}}{dr}-q_{nl}\frac{dp_{nl}}{dr} \ = \ -\frac{1}{r^{2}f} \ . 
\end{equation}
The ordinary WKB forms are given if we assume $p_{nl}$ and $q_{nl}$ to be proportional to $I_{1/2}$ and $K_{1/2}$, respectively. Now, the function $w$ introduced in place of $W$ should satisfy the equation 
\begin{eqnarray}
\frac{w^{2}}{f^{2}}\{1+\frac{1}{\chi^{2}}(n^{2}-\frac{1}{4})\} \ = \ \frac{n^{2}\kappa^{2}}{f^{2}}+\frac{l(l+1)+m^{2}r^{2}}{fr^{2}} \nonumber \\
+\frac{1}{2w}\frac{d^{2}w}{dr^{2}}-\frac{3}{4}\frac{1}{w^{2}}(\frac{dw}{dr})^{2}+\frac{1}{2r^{2}fw}\frac{d(r^{2}w)}{dr}\frac{df}{dr} \ . 
\label{weq}
\end{eqnarray}
If $w$ is rewritten into
\begin{equation} 
w \ \equiv \ f^{1/2}y/r_{+} \ ,
\end{equation} 
the solution of Eq.\ (\ref{weq}) allows the expansion form
\begin{equation}
y \ = \ B(1+\sum_{s=1}^{\infty}y_{s}z^{s}) \ . \label{yexp}
\end{equation}
From the well-known behavior of the modified Bessel function $K_{n}(\chi)$ near $\chi=0$, it is easy to see that $q_{nl}$ has the expected logarithmic behavior near the event horizon. 

By substituting Eq.\ (\ref{yexp}) into Eq.\ (\ref{weq}) with the expansion in powers of $z$, we obtain the recurrence relation between the coefficients $B$ and $y_{s}$. For example, the lowest relation leads to 
\begin{equation}
\frac{2\kappa r_{+}}{3}(n^{2}-1)(y_{1}-2+\frac{1}{2\kappa r_{+}}) \ = \ \nu(\nu+1)+2\kappa r_{+}-B^{2} \ , \label{lowest} 
\end{equation} 
where $\nu(\nu+1)=l(l+1)+m^{2}r_{+}^{2}$. From the expansion up to the next power of $z$ the relation between $y_{1}$ and $y_{2}$ turns out to be 
\begin{equation}
\frac{2\kappa r_{+}}{5}(n^{2}-4)y_{2} \ = \ -\nu(\nu+1)y_{1}-l(l+1)+U(\kappa r_{+},n,y_{1}) \ , \label{next}
\end{equation}
where $U$ is a slightly complicated quadratic function of $y_{1}$ which depends on $n$ and $\kappa r_{+}$ only. An important point of the expansion form (\ref{yexp}) is that we can require $y_{s}$ to remain finite in the limit $l\rightarrow\infty$, for which from Eqs.\ (\ref{lowest}) and (\ref{next}) the asymptotic values of $B$ and $y_{1}$ reduce to 
\begin{equation}
B^{2} \ = \ l(l+1)+m^{2}r_{+}^{2}+\frac{1}{3}+n^{2}(2\kappa r_{+}-\frac{1}{3})+O(l^{-2}) \ , \label{asB}
\end{equation}
and
\begin{equation}
y_{1} \ = \ -1+O(l^{-2}) \ , 
\end{equation}
This dependence of $y_{s}$ on $l$ allows us to calculate the mode sum over $l$ in $A_{n}$ by neglecting the terms with the higher powers of $z$ in Eq.\ (\ref{yexp}), and in the following Eq.\ (\ref{asB}) will be verified in terms of the cancellation of the logarithmic divergence in $<\phi^{2}>_{ren}$. 

We also remark that the amplitude of $<\phi^{2}>_{ren}$ at the event horizon should not be interpreted as a quantity determined only by local geometry. The relations (\ref{lowest}) and (\ref{next}) allow us to give a conjecture that the recurrence relation is truncated within a finite sequnce, and for the $n$-th mode the finite set consisted of $B$, $y_{1}$, $\cdots$, $y_{n-1}$ is completely determined for any value of $l$. However, the coefficient $y_{n}$ remains unknown, unless the higher infinite sequnce of the recurrence relation is consistently solved for satisfying the boundary condition $y\rightarrow (m^{2}r_{+}^{2}+n^{2}\kappa^{2}r_{+}^{2})^{1/2}$ at $z\rightarrow\infty$ as an eigenvalue problem. In particular, for $n=0$ we cannot give $B$ for lower values of $l$ without a further analysis of Eq.\ (\ref{qnl}). This is the problem to be solved in the subsequent sections, and in this section we use Eq.\ (\ref{asB}) for $n=0$ to derive the logarithmic term in $A_{0}$.

By taking the limit $z\rightarrow0$, we can give the mode sum over $l$ for $n=0$ written by the form 
\begin{equation}
A_{0} \ = \ \sum_{l=0}^{\infty}\{\frac{2l+1}{\kappa r_{+}^{2}}K_{0}(B\sqrt{2z/\kappa r_{+}})I_{0}(B\sqrt{2z/\kappa r_{+}})-\frac{1}{r_{+}\sqrt{2\kappa r_{+}z}}\} . \label{A0}
\end{equation}
Then, we apply the Plana sum formula (\ref{Plana}) to Eq.\ (\ref{A0}), in which the modified Bessel functions is allowed to reduce to 
\begin{equation}
K_{0}(B\sqrt{2z/\kappa r_{+}}) \ \simeq \ -\gamma-\ln(B\sqrt{z/2\kappa r_{+}}) \ , \label{K0}
\end{equation}
and
\begin{equation}
I_{0}(B\sqrt{2z/\kappa r_{+}}) \ \simeq \ 1 \ , \label{I0}
\end{equation}
except for the integral defined by
\begin{equation}
\int_{0}^{\infty}dl\{\frac{2l+1}{\kappa r_{+}^{2}}K_{0}(B\sqrt{2z/\kappa r_{+}})I_{0}(B\sqrt{2z/\kappa r_{+}})-\frac{1}{r_{+}\sqrt{2\kappa r_{+}z}}\} \ . \label{integral}
\end{equation}
To calculate the integral (\ref{integral}), let us recall that $B$ is a function of $l$ satisfying 
\begin{equation}
2BdB/dl \ = \ 2l+1+O(l^{-2})
\end{equation}
in the large $l$ limit and replace the integral of the modified Bessel functions over $l$ by that over $B$ to use the integral formula 
\begin{equation}
\int 2BK_{0}(Bv)I_{0}(Bv)dB \ = \ B^{2}\{K_{0}(Bv)I_{0}(Bv)+K_{1}(Bv)I_{1}(Bv)\} 
\end{equation}
for any variable $v$. Then, the same approximations with Eqs.\ (\ref{K0}) and (\ref{I0}) is applicable to the remaining integral given by
\begin{equation}
\int_{0}^{\infty}\frac{dl}{\kappa r_{+}^{2}}(2l+1-2B\frac{dB}{dl})K_{0}(B\sqrt{2z/\kappa r_{+}})I_{0}(B\sqrt{2z/\kappa r_{+}}) \ , 
\end{equation}
and we arrive at the final result for $A_{0}$ in the limit $z\rightarrow0$ such that
\begin{equation}
A_{0} \ = \ \frac{S_{0}}{\kappa r_{+}^{2}}+\frac{m^{2}}{\kappa}\{\gamma+\frac{1}{2}\ln(\frac{z}{2\kappa r_{+}})\} \ , 
\end{equation}
where 
\begin{eqnarray}
S_{0} \ = \ (B_{0}^{2}-\frac{1}{2})\ln B_{0}-\frac{B_{0}^{2}}{2}-\int_{0}^{\infty}dl(2l+1-2B\frac{dB}{dl})\ln B \nonumber \\
-\int_{0}^{\infty}\frac{idl}{e^{2\pi l}-1}\{(2il+1)\ln B(il)+(2il-1)\ln B(-il)\} \ , \label{S0}
\end{eqnarray}
if we denote $B(l=0)$ by $B_{0}$. Hence, by adding $\kappa A_{0}/8\pi^{2}$ to $<\phi^{2}>_{ap}$, the logarithmic divergence at the event horizon turns out to be canceled, and we obtain the renormalized value denoted by $<\phi^{2}>_{H}$ as follows, 
\begin{equation}
<\phi^{2}>_{H} \ = \ \frac{\kappa}{24\pi^{2}r_{+}}+\frac{m^{2}}{16\pi^{2}}\{1-\ln(m^{2}r_{+}^{2}))\}+\frac{S_{0}}{8\pi^{2}r_{+}^{2}} \ . \label{horizon}
\end{equation}

It is interesting to note that the absence of the logarithmic divergence of $<\phi^{2}>_{ren}$ at the event horizon is assured only by giving the asymptotic value (\ref{asB}) of $B$ for the $n=0$ mode with very large $l$, which is determined through the local analysis near $r=r_{+}$. Though in general we cannot obtain the renormalized value itself without deriving $B$ for lower $l$ modes, the large-mass limit can be an exceptional case for which the local analysis remains useful, and we calculate $<\phi^{2}>_{H}$ up to the order of $m^{-2}$ in the next section as a simple application of the procedure presented here.

\section{the large-mass limit}
\label{largemass}
To calculate the integral of $B$ in $S_{0}$ over $l$ under the large-mass limit $mr_{+}\gg1$, it is convenient to give the expansion form of $B$ in inverse powers of $\nu(\nu+1)$, by keeping the quantity $\mu\equiv m^{2}r_{+}^{2}/\nu(\nu+1)$ to be of order of unity. (For the first integral present in $S_{0}$ we cannot assume $l(l+1)$ to be much smaller than $mr_{+}$, while for the second integral the approximation 
$\mu\simeq 1-l(l+1)(mr_{+})^{-2}$ may be allowed.) The expansion of $B^{2}$ should be done up to the terms of order of $1/\nu(\nu+1)$ for obtaining the $m^{-2}$ terms of $<\phi^{2}>_{H}$. Then, the recurrence relation subsequent to Eqs.\ (\ref{lowest}) and (\ref{next}) becomes necessary, for which the leading terms turn out to be
\begin{equation}
y_{2} \ = \ -\frac{y_{1}^{2}}{2}+\frac{3}{2}(1-\mu)+O(m^{-2}) \ . \label{3rd}
\end{equation}
The key point of Eq.\ (\ref{3rd}) is the absence of $y_{3}$ in the leading-order relation, from which Eqs.\ (\ref{lowest}) and (\ref{next}) for $n=0$ can give
\begin{equation}
y_{1} \ = \ -1+\mu+\frac{\kappa r_{+}}{\nu(\nu+1)}\eta+0(m^{-4}) \ , 
\end{equation}
and
\begin{equation}
B^{2} \ = \ \nu(\nu+1)+\frac{1}{3}(1+2\kappa r_{+}\mu)+\frac{2\kappa^{2}r_{+}^{2}}{3\nu(\nu+1)}\eta+O(m^{-4}) \ , 
\end{equation}
where
\begin{equation}
\eta \ = \ -\frac{1}{60\kappa^{2}r_{+}^{2}}+(\frac{4}{5}-\frac{1}{15\kappa r_{+}})\mu-\frac{37}{15}\mu^{2} \ . 
\end{equation}

Now it is easy to calculate the integrals in Eq.\ (\ref{S0}) up to the terms of order of $(mr_{+})^{-2}$, and we can confirm the cancellation of all the terms much larger than $(mr_{+})^{-2}$ in the expression (\ref{horizon}) for $<\phi^{2}>_{H}$, giving the result
\begin{equation}
<\phi^{2}>_{H} \ = \ \frac{1}{720\pi^{2}m^{2}r_{+}^{4}}(16\kappa^{2}r_{+}^{2}-4\kappa r_{+}+1) \ , \label{m2}
\end{equation}
Note that the well-known $m^{-2}$ term $<\phi^{2}>_{m^{-2}}$ of the DeWitt-Schwinger approximation for $<\phi^{2}>$ can be written by
\begin{equation}
<\phi^{2}>_{m^{-2}} \ = \ \frac{1}{2880\pi^{2}m^{2}}(R_{abcd}R^{abcd}-R_{ab}R^{ab}) 
\end{equation}
for Reissner-Nordstr\"{o}m background (with vanishing Ricci scalar), where $R_{abcd}$ and $R_{ab}$ are the Riemann and Ricci tensors, respectively. If evaluated at the event horizon $r=r_{+}$, this DeWitt-Schwinger term is found to be identical with Eq.\ (\ref{m2}). Hence, for very massive fields with $mr_{+}\gg1$ in equilibrium at black-hole temperature $T=\kappa/2\pi$, we can claim the validity of the DeWitt-Schwinger approximation near the event horizon, as was previously shown in numerical calculations \cite{An95,An90}. Further, if $mr_{+}$ is fixed, the tail part (\ref{m2}) in the range $mr_{+}\gg1$ becomes minimum at the black-hole temperature corresponding to $\kappa r_{+}=1/8$, rather than at the low-temperature limit $\kappa r_{+}\ll1$. The $m$-$\kappa$ coupling can give a slightly complicated change to the amplitude of vacuum polarization. In the next section we see a result of the $m$-$\kappa$ coupling as the suppression of temperature-induced excitation in a small-mass range.  

\section{the small-mass limit}
\label{smallmass}
Now we consider scalar fields with very small mass $mr_{+}\ll1$, for which the temperature-induced excitaion given by Eq.\ (\ref{massless}) will dominate. To reveal some correction due to the small field mass, let us begin with a brief analysis of purely massless fields. It is easy to see that Eq.\ (\ref{qnl}) for the massless $n=0$ modes becomes equal to Legendre's differential equation, if we use the variable $x$ defined by $x \ = \ 1+(z/\kappa r_{+})$. Then, from the behavior of Legendre functions at $x\rightarrow1$ and $x\rightarrow\infty$, the modes $q_{0l}$ and $p_{0l}$ should be chosen to be 
\begin{equation}
q_{0l} \ = \ Q_{l}(x) \ , \ \ \ p_{0l} \ = \ P_{l}(x) \ ,
\end{equation}
The mode sum in Eq.\ (\ref{sum}) for $n=0$ is known to be precisely zero for any $x$ \cite{C-H}, and from Eq.\ (\ref{horizon}) the vacuum polarization at the event horizon reduces to 
\begin{equation}
<\phi^{2}>_{H} \ = \ \frac{\kappa}{24\pi^{2}r_{+}} \ , \label{thermal}
\end{equation}
which should be interpreted to be purely induced by the black-hole temperature. For purpose of extending the result to massive fields, it is useful to check explicitly through the procedure given in the previous sections that $S_{0}$ in Eq.\ (\ref{horizon}) vanishes.

Recall that the function $Q_{l}(x)$ has logarithmic branch point at $x=1$, and the dominant behavior near the point is
\begin{equation}
Q_{l} \ \simeq \ \frac{1}{2}\ln(\frac{2}{x-1})-\psi(1+l)-\gamma \ , 
\end{equation}
where $\psi(s)$ is the logarithmic derivative of the gamma function (i.e., a polygamma function), and we have $\psi(1)=-\gamma$ for Euler's constant $\gamma$. By comparing the logarithmic behavior of $Q_{l}$ with Eq.\ (\ref{K0}) for the modified Bessel function, we can determine the coefficient $B$ as follows,
\begin{equation}
B \ = \ \exp\{\psi(1+l)\} \ .
\end{equation} 
To calculate the integrals over $l$ in $S_{0}$, we use integral representations for the polygamma function. For example, we obtain
\begin{eqnarray}
-\int_{0}^{\infty}\frac{idl}{e^{2\pi l}-1}\{(2il+1)\psi(1+il)+(2il-1)\psi(1-il)\} \ = \nonumber \\ 
\int_{0}^{\infty} dt\{\frac{e^{-t}}{6t}-\frac{2t^{-2}+t^{-1}}{e^{t}-1}+\frac{1}{4}(\frac{\cosh(t/2)}{\sinh^{3}(t/2)}-\coth(t/2)+1)\} \ , \label{int1}
\end{eqnarray}
by virtue of the formula  
\begin{equation}
\psi(s) \ = \ \int_{0}^{\infty}dt(\frac{e^{-t}}{t}-\frac{e^{-ts}}{1-e^{-t}}) \ . \label{int2}
\end{equation}
Another useful formula is given by
\begin{equation}
\psi(s) \ = \ \ln s-\frac{1}{2s}-\frac{1}{12s^{2}}-\int_{0}^{\infty} dt(\frac{1}{e^{t}-1}-\frac{1}{t}+\frac{1}{2}-\frac{t}{12})e^{-ts} \ , \label{int3}
\end{equation}
through which we arrive at the result
\begin{eqnarray}
\int_{0}^{\infty} dl\{2e^{2\psi(1+l)}\frac{d\psi(1+l)}{dl}-(2l+1)\}\psi(1+l) \ = \nonumber \\ 
(\frac{1}{2}+\gamma)e^{-2\gamma}-\frac{1}{3}+\int_{0}^{\infty} dt(\frac{1}{e^{t}-1}-\frac{1}{t}+\frac{1}{2}-\frac{t}{12})(\frac{2}{t^{2}}+\frac{1}{t}) \ . \label{int4}
\end{eqnarray}
Then, it becomes easy to calculate the integral over $t$ for the sum of Eqs.\ (\ref{int1}) and (\ref{int4}), and we obtain $S_{0}=0$.

For the massive $n=0$ mode we rewrite Eq.\ (\ref{qnl}) into the form
\begin{equation}
(x^{2}-1)\frac{d^{2}q_{0l}}{dx^{2}}+2x\frac{dq_{0l}}{dx}-\{l(l+1)+m^{2}r_{+}^{2}(\kappa r_{+}x+1-\kappa r_{+})^{2}\}q_{0l} \ = \ 0 \ , \label{plegendre}
\end{equation}
which can clarify the deviation from Legendre's differential equation. In this section a small-mass field having $mr_{+}\ll1$ is assumed, and the solution perturbed by the field mass is given by 
\begin{equation}
q_{0l} \ = \ Q_{l'}(x)+q_{l}(x) \ , 
\end{equation}
where $l'-l\equiv\delta=O(m^{2}r_{+}^{2})$. Because the terms proportional to $m^{2}r_{+}^{2}$ in Eq.\ (\ref{plegendre}) is dependent on $x$, we use the recurrence formula valid for $Q_{l'}$ (and also for $P_{l'}$) such that
\begin{equation}
(l'+1)Q_{l'+1}-(2l'+1)xQ_{l'}+l'Q_{l'-1} \ = \ 0 \ , 
\end{equation}
and the perturbed part $q_{l}$ is expanded in terms of Legendre functions as follows,
\begin{equation}
q_{l} \ = \ \sum_{k=1}^{\infty}(c_{k}^{(l)}Q_{l'+k}+c_{-k}^{(l)}Q_{l'-k}) \ . 
\end{equation}
The coefficients $c_{k}$ and $c_{-k}$ together with the eigenvalue $\delta$ are determined by solving the recurrence relation
\begin{equation}
c_{k}^{(l)}\{(l'+k)(l'+k+1)-l(l+1)-m^{2}r_{+}^{2}v_{l'+k}^{(0)}\} \ = \ m^{2}r_{+}^{2}\sum_{j=1}^{2}(v_{l'+k}^{(j)}c_{k+j}^{(l)}+v_{l'+k}^{(-j)}c_{k-j}^{(l)}) \ , 
\end{equation}
where $c_{0}^{(l)}=1$, and 
\begin{eqnarray}
v_{i}^{(0)} \ = \ (1-\kappa r_{+})^{2}+\kappa^{2}r_{+}^{2}\frac{2i(2i+1)-1}{(2i-1)(2i+3)} \ , \nonumber \\
v_{i}^{(1)} \ = \ 2\kappa r_{+}(1-\kappa r_{+})\frac{i+1}{2i+3} \ , \ \ v_{i}^{(-1)} \ = \ 2\kappa r_{+}(1-\kappa r_{+})\frac{i}{2i-1} \ , \\
v_{i}^{(2)} \ = \ \kappa^{2}r_{+}^{2}\frac{(i+1)(i+2)}{(2i+3)(2i+5)} \ , \ \ v_{i}^{(-2)} \ = \ \kappa^{2}r_{+}^{2}\frac{i(i-1)}{(2i-3)(2i-1)} \ . \nonumber
\end{eqnarray}

Then, the first-order perturbation is found to be 
\begin{equation}
q_{l} \ = \ \frac{m^{2}\kappa r_{+}^{3}}{2l+1}\{(1-\kappa r_{+})(Q_{l+1}-Q_{l-1})+\frac{\kappa r_{+}}{2}(\frac{(l+1)(l+2)Q_{l+2}}{(2l+3)^{2}}-\frac{l(l-1)Q_{l-2}}{(2l-1)^{2}})\} \ , \label{deltaq}
\end{equation}
and
\begin{equation}
\delta \ = \ \frac{m^{2}r_{+}^{2}}{2l+1}\{(1-\kappa r_{+})^{2}+\kappa^{2}r_{+}^{2}\frac{2l(l+1)-1}{(2l-1)(2l+3)}\} \ ,  
\end{equation}
for which the coefficient $B$ is estimated to be
\begin{equation}
B \ = \ e^{\psi(l+1)}\{1+\delta\frac{d\psi(l+1)}{dl}+m^{2}r_{+}^{2}(\frac{\kappa r_{+}(1-\kappa r_{+})}{l(l+1)}+\frac{\kappa^{2}r_{+}^{2}}{(2l-1)(2l+3)})\} \ , 
\end{equation} 
Using these equations, one may calculate the polarization amplitude $<\phi^{2}>_{H}$ at the event horizon. However, for $l=0$ the value of $B$ becomes divergent as a result of the existence of the undefined function $Q_{-k}$ in Eq.\ (\ref{deltaq}). This will mean a dominant contribution of the $l=0$ mode in the small-mass limit. 

To estimate more precisely $B=B_{0}$ for $l=0$, the subscript $l$ in the Legendre functions should be replaced by $l'$, taking account of the approximate relation $Q_{\delta-k}\simeq P_{k-1}/\delta$ for $\delta\ll1$. Then, the term $m^{2}r_{+}^{2}Q_{\delta-1}$ which appears in $q_{0}$ should be interpreted to be of order of unity, contradictory to the perturbation scheme. This problem is resolved if we add another independent solution for Eq.\ (\ref{plegendre}) written by
\begin{equation}
p_{0} \ = \ d_{0}^{(0)}P_{\delta}+\sum_{k=1}^{\infty}(d_{k}^{(0)}P_{\delta+k}+d_{-k}^{(0)}P_{\delta-k}) 
\end{equation}
to $q_{0}$ as follows,
\begin{equation}
q_{0} \ = \ \sum_{k=1}^{\infty}(c_{k}^{(0)}Q_{\delta+k}+c_{-k}^{(0)}Q_{\delta-k})+p_{0} \ , 
\end{equation}
where we require that $\delta^{-1}c_{-1}^{(0)}+d_{0}^{(0)}\equiv\varepsilon\ll1$ for $d_{0}^{(0)}$ of order of unity. Of course, the coefficients $d_{k}^{(0)}$ should satisfy the same recurrence relation with $c_{k}^{(0)}$, and we obtain for $k\geq1$
\begin{equation}
d_{2k-1}^{(0)} \ = \ O((mr_{+})^{2k}) \ , \ \ d_{2k}^{(0)} \ = \ O((mr_{+})^{2k}) \ , 
\end{equation}
in addition to the ratio $d_{-k}^{(0)}/d_{k-1}^{(0)}=O(m^{2}r_{+}^{2})$. Then, the asymptotic behavior of the $l=0$ mode $q_{00}$ at $x\gg1$ is approximately given by
\begin{equation}
q_{00} \ \simeq \ \frac{1}{x}+\sum_{k=0}^{\infty} \frac{\Gamma(k+(1/2))}{\sqrt{\pi}\Gamma(k+1)}(\delta^{-1}c_{-k-1}^{(0)}+d_{k}^{(0)})(2x)^{k} 
\ , 
\end{equation}
which should be consistent with the boundary condition
\begin{equation}
q_{00} \simeq \frac{1}{x}\exp(-m\kappa r_{+}^{2}x)
\end{equation}
at a distant region far from the event horizon. 

To check the consistency, let us derive the approximate recurrence relation which is valid up to the leading order of $m^{2}r_{+}^{2}$ and reduces to 
\begin{equation}
\frac{c_{-1-2k}^{(0)}}{c_{1-2k}^{(0)}} \ = \ \frac{d_{2k}^{(0)}}{d_{2k-2}^{(0)}} \ = \ m^{2}\kappa^{2}r_{+}^{4}\frac{2k-1}{(2k+1)(4k-1)(4k-3)} \ ,
\end{equation}
and
\begin{equation}
\frac{\delta^{-1}c_{-2k-2}^{(0)}+d_{2k+1}^{(0)}}{\delta^{-1}c_{-2k}^{(0)}+d_{2k-1}^{(0)}} \ = \ m^{2}\kappa^{2}r_{+}^{4}\frac{2k}{(2k+2)(4k+1)(4k-1)} \ . 
\end{equation}
Noting the relations between the lowest coefficients such that
\begin{equation}
\delta^{-1}c_{-1}^{(0)} \ = \ 2m^{2}\kappa r_{+}^{3}(1-\kappa r_{+}) 
\end{equation}
and 
\begin{equation}
\delta^{-1}c_{-2}^{(0)}+d_{1}^{(0)}=m^{2}\kappa^{2}r_{+}^{4}/2 \ ,
\end{equation}
we arrive at the result
\begin{equation}
q_{00} \ \simeq \ \sum_{k=1}^{\infty}\frac{(m\kappa r_{+}^{2}x)^{2k}}{x(2k)!}+\varepsilon\sum_{k=1}^{\infty}\frac{(m\kappa r_{+}^{2}x)^{2k-2}}{(2k-1)!} \ , 
\end{equation}
which can satisfy the boundary condition if $\varepsilon=-m\kappa r_{+}^{2}$. 

Unfortunately, we cannot determine $\varepsilon$ to the order of $m^{2}r_{+}^{2}$, unless the reccurence relation is studied to the higher order. Hence, we only keep the leading correction of order of $mr_{+}$ in the $l=0$ mode, 
\begin{equation}
q_{00} \ \simeq \ Q_{0}-m\kappa r_{+}^{2} \ , 
\end{equation}
which means that $B_{0}=e^{-\gamma}(1+m\kappa r_{+})$. For the $l\geq1$ modes $q_{0l}$ we must also consider the perturbation with the terms written by the Legendre functions $P_{k}(x)$. However, it is sure that no perturbation of order of $mr_{+}$ does not appear for $l\geq1$, and we obtain\begin{equation}
S_{0} \ \simeq \ -\ln(1+m\kappa r_{+}^{2}) \ \simeq \ -m\kappa r_{+}^{2} \ , 
\end{equation}
if we omit the higher-order corrections. Now the vacuum polarization given by Eq.\ (\ref{horizon}) for small-mass fields becomes approximately 
\begin{equation}
<\phi^{2}>_{H} \ \simeq \ \frac{\kappa}{24\pi^{2}r_{+}}(1-3mr_{+}) \ , 
\end{equation}
which cleary shows that the temperature-induced excitation is suppressed by field mass. As $m$ becomes larger, the amplitude may monotoneously decrease in the whole mass range extending to $mr_{+}\gg1$ where the DeWitt-Schwinger approximation $<\phi^{2}>_{H} \sim (mr_{+})^{-2}$ is valid. This simple dependence on $m$ is supported through numerical calculations for several values of $mr_{+}$ in Schwarzschild background ($\kappa r_{+}=1/2$) \cite{An89}. In the next section, however, we point out a different dependence on field mass, which is a resonant behavior of $<\phi^{2}>_{H}$ remarkable in the low-temperature case $\kappa r_{+}\ll1$.

\section{mass-induced excitation}
\label{mass-induced}
Let us turn attention to quantum fields at the event horizon of nearly extreme black holes to show an interesting feature of the mass-induced excitation of vacuum polarization. Then, we do not limit the range of the parameter $mr_{+}$, but we solve Eq.\ (\ref{plegendre}) under the assumption $\kappa r_{+}\ll1$ by the help of the technique of asymptotic matching. 

At large values of $x$ Eq.\ (\ref{plegendre}) reduces to the form
\begin{equation}
\frac{d^{2}q_{0l}}{dx^{2}}+\frac{2}{x}\frac{dq_{0l}}{dx}-(\frac{\nu(\nu+1)}{x^{2}}+\frac{2m^{2}\kappa r_{+}^{3}}{x}+m^{2}\kappa^{2}r_{+}^{4})q_{0l} \ = \ 0 \ , 
\end{equation}
in which we cannot neglect the terms depending on $\kappa r_{+}$ to require the exponential decrease of $q_{0l}$. For the approximate differential equation we obtain the solution
\begin{equation}
q_{0l} \ = \ W_{-mr_{+},\nu+\frac{1}{2}}(2m\kappa r_{+}^{2}x)/x \ , 
\end{equation}
where $W_{a,b}$ denotes the Whittaker function with the asymtotic behavior
\begin{equation}
W_{a,b}(u) \ \simeq \ u^{a}\exp(-u/2) 
\end{equation}
as $u\rightarrow\infty$. This asymptotic soluition can remain valid in the range 
\begin{equation}
1 \ \ll \ x \ \ll \ 1/\kappa r_{+} \ , 
\end{equation}
where we obtain the approximate behavior
\begin{equation}
q_{0l} \ \simeq \ \frac{\Gamma(-2\nu-1)}{\Gamma(mr_{+}-\nu)}(2m\kappa r_{+}^{2}x)^{\nu+1}x^{-1}+\frac{\Gamma(2\nu+1)}{\Gamma(mr_{+}+\nu+1)}(2m\kappa r_{+}^{2}x)^{-\nu}x^{-1} \ . \label{matching}
\end{equation}
Note that if $x\ll1/\kappa r_{+}$, Eq.\ (\ref{plegendre}) becomes approximately equal to Legendre's differential equation, giving the solution
\begin{equation}
q_{0l} \ = \ CP_{\nu}(x)+DQ_{\nu}(x) \ .
\end{equation}
The coefficients $C$ and $D$ should be determined by the matching with the approximate solution (\ref{matching}), and it is easy to see that the ratio $C/D$ is of order of $(m\kappa r_{+}^{2})^{2\nu+1}$. Hence, we can neglect the term $P_{\nu}$ in $q_{0l}$, and the asymtotic behavior at $x\rightarrow1$ turns out to be
\begin{equation}
q_{0l} \ \simeq \ -D\{\frac{1}{2}\ln(\frac{x-1}{2})+\gamma+\psi(\nu+1)\} \ , 
\end{equation}
from which we obtain
\begin{equation}
B \ = \ e^{\psi(\nu+1)} \ , 
\end{equation}
for calculating $S_{0}$ (and $<\phi^{2}>_{H}$) through Eq.\ (\ref{S0}). 

A useful expression of $S_{0}$ to understand the field-mass dependence is derived if we use the integral formula
\begin{equation}
\psi(\nu+1) \ = \ \frac{1}{2}\ln(\nu^{2}+\nu+\frac{1}{4})+\int_{0}^{\infty}\frac{2tdt}{(e^{2\pi t}+1)(t^{2}+\nu^{2}+\nu+(1/4))} \ . 
\end{equation}
In fact, for $F(l)\equiv(-i)\{(2il+1)\ln B(il)+(2il-1)\ln B(-il)\}$ which is one of the integrands in $S_{0}$, we obtain 
\begin{equation}
F(l) \ = \ l\ln\{(l^{2}-\zeta)^{2}+l^{2}\}+\arctan(\frac{l}{\zeta-l^{2}})-\int_{0}^{\infty}\frac{8tdt}{e^{2\pi t}+1}\frac{l^{2}+(1/2)-t^{2}-\zeta}{(l^{2}-t^{2}-\zeta)^{2}+l^{2}} \ , 
\end{equation}
where $\zeta=m^{2}r_{+}^{2}+(1/4)$, and the value of $\arctan(u)$ runs from $0$ to $\pi$ in the range $0\leq u\leq\infty$. Further, the integral given by 
\begin{equation}
\int \ dl(2l+1-2B\frac{dB}{dl})\ln B 
\end{equation}
is rewritten into the form
\begin{eqnarray}
\frac{1}{2}\{\nu(\nu+1)+\frac{1}{4}\}\{\ln(\nu(\nu+1)+\frac{1}{4})-1\}-e^{2\psi(\nu+1)}\{\psi(\nu+1)-\frac{1}{2}\} \nonumber \\
+2\int_{0}^{\infty}\frac{tdt}{e^{2\pi t}+1}\ln(t^{2}+\nu(\nu+1)+\frac{1}{4}) \ ,
\end{eqnarray}
which is equal to zero as $l\rightarrow\infty$. We therefore arrive at the result
\begin{equation}
S_{0} \ = \ \frac{1}{2}(\zeta-\frac{1}{2})\ln\zeta-\frac{\zeta}{2}+\int_{0}^{\infty}\{\frac{tG(t)}{e^{2\pi t}+1}+\frac{H(t)}{e^{2\pi t}-1}\}dt
\end{equation}
where
\begin{equation}
G(t) \ = \ 2\ln(t^{2}+\zeta)-\frac{1}{t^{2}+\zeta}-8\int_{0}^{\infty}\frac{dl}{e^{2\pi l}-1}\frac{l^{2}+(1/2)-t^{2}-\zeta}{(l^{2}-t^{2}-\zeta)^{2}+l^{2}} \ , 
\end{equation}
and
\begin{equation}
H(t) \ = \ t\ln\{(t^{2}-\zeta)^{2}+t^{2}\}+\arctan(\frac{t}{\zeta-t^{2}}) \ .
\end{equation}
Under the low-temperature approximation $\kappa r_{+}\ll1$ we neglect the term $\kappa/24\pi^{2}r_{+}$ in Eq.\ (\ref{horizon}), and the polarization amplitude at the event horizon is finally given by
\begin{equation}
8\pi^{2}r_{+}^{2}<\phi^{2}>_{H} \ = \ \frac{m^{2}r_{+}^{2}}{2}\ln(\frac{\zeta}{m^{2}r_{+}^{2}})-\frac{1}{8}(1+\ln\zeta)+\int_{0}^{\infty}\{\frac{tG(t)}{e^{2\pi t}+1}+\frac{H(t)}{e^{2\pi t}-1}\}dt \ .
\end{equation}

Now it is easy to check the value of $<\phi^{2}>_{H}$ in the large-mass limit $mr_{+}\gg1$, and we obtain
\begin{equation}
8\pi^{2}r_{+}^{2}<\phi^{2}>_{H} \ \simeq \ \frac{1}{90m^{2}r_{+}^{2}} \ , \label{tail}
\end{equation}
for which we can reconfirm that it is equal to the DeWitt-Schwinger approximation (with $\kappa r_{+}\rightarrow0$). We can also consider the small-mass limit $mr_{+}\ll1$ under the condition $m/\kappa\gg1$, and the approximate expression of $<\phi^{2}>_{H}$ becomes
\begin{equation}
8\pi^{2}r_{+}^{2}<\phi^{2}>_{H} \ \simeq \ -m^{2}r_{+}^{2}\{\frac{1}{2}+\gamma+\ln(mr_{+})\} \ , 
\end{equation}
which can remain positive by virtue of the existence of the logarithmic term $-m^{2}r_{+}^{2}\ln(mr_{+})$. 

\begin{figure}
\begin{center}
\epsfxsize=150mm
\epsfbox{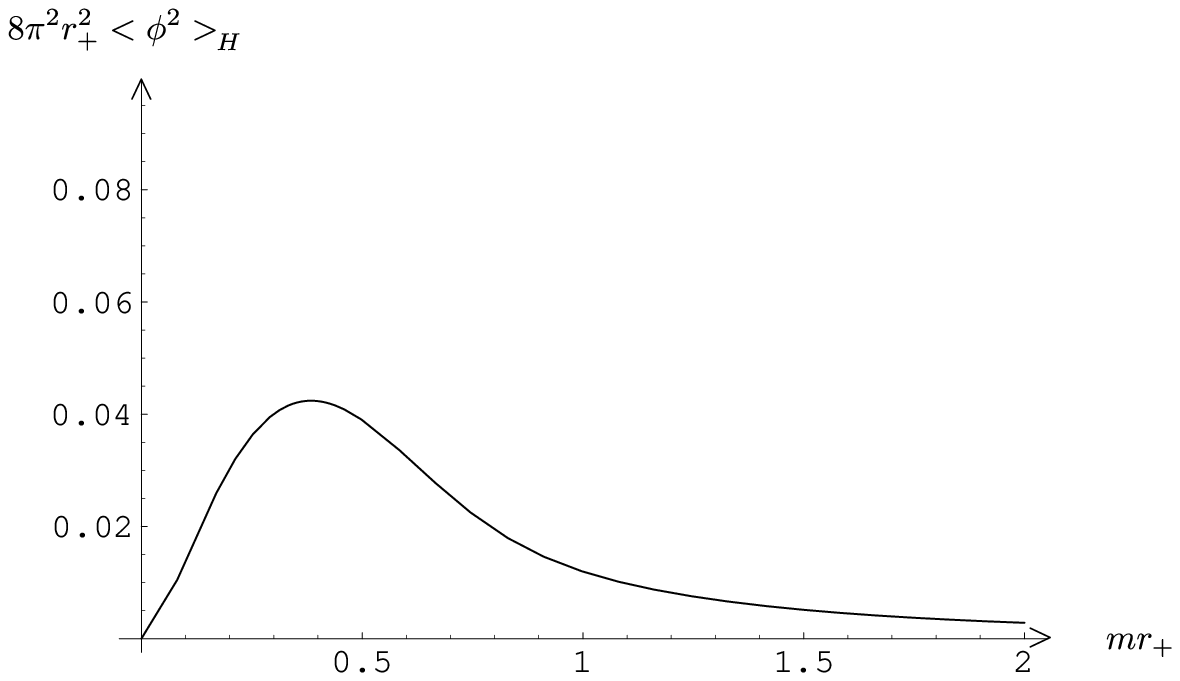}         
\caption{The field-mass dependence of vacuum polarization $<\phi^{2}>_{H}$ at the nearly extreme Reissner-Nordstr\"{o}m horizon $r=r_{+}$. The amplitude has a resonance peak at $mr_{+}\simeq 0.38$ and a tail part decreasing in proportion to $m^{-2}$ for very massive fields.}
\label{fig1}
\end{center}
\end{figure}

We evaluate numerically the integrals in the expression of $<\phi^{2}>_{H}$, and the field-mass dependence is shown in Fig.\ \ref{fig1}. Note that the maximum excitation of $<\phi^{2}>_{H}$ occurs at $mr_{+}\simeq 0.38$, and the peak amplitude denoted by $<\phi^{2}>_{max}$ is estimated to be $8\pi^{2}r_{+}^{2}<\phi^{2}>_{max}\simeq 0.0424$. We can clearly see a resonance behavior of the polarization amplitude for massive fields with the Compton wavelength $1/m$ of order of $r_{+}$ and also the tail part given by Eq.\ (\ref{tail}) in the mass range of $mr_{+}\gg1$.

\section{summary}
\label{summary}
We have studied vacuum polarization of quantized scalar fields in Reissner-Nordstr\"{o}m background by means of the Euclidean space Green's function. In particular, the renormalized expression $<\phi^{2}>_{H}$ at the event horizon $r=r_{+}$ has been derived by revealing the contribution of the $n=0$ mode, which can cancel the logarithmic divergence.

We have found the dependence of $<\phi^{2}>_{H}$ on field mass $m$: (1) The tail part observed in the large-mass limit $mr_{+}\gg1$ becomes equal to the DeWitt-Schwinger approximation. (2) For small-mass fields a suppression of temperature-induced excitation due to the coupling between $m$ and $\kappa$ occurs according to $<\phi^{2}>_{H}=<\phi^{2}>_{T}(1-3mr_{+})$, where the massless part with the amplitude proportional to the black-hole temperature $T=\kappa/2\pi$ is given by $8\pi^{2}r_{+}^{2}<\phi^{2}>_{T}=\kappa r_{+}/3$. We can expect that mass-induced excitation becomes important for massive fields with $mr_{+}\simeq1$. Unfortunately, it is difficult to investigate in detail various aspects of the $m$-$\kappa$ coupling in the case that both $mr_{+}$ and $\kappa r_{+}$ are of order of unity. (3) Our main result therefore has been to show a resonance behavior of mass-induced excitation of vacuum polarization around nearly extreme Reissner-Nordstr\"{o}m black holes with $\kappa r_{+}\ll1$: If the Compton wavelength $1/m$ of a massive field is of order of the black-hole radius $r_{+}$, the amplitude of vacuum polarization has a peak at the resonance mass given by $mr_{+}\simeq 0.38$. 

There should be a critical temperature $T_{c}=\kappa_{c}/2\pi$ of black holes in the range $0<\kappa r_{+}<1/2$, below which a resonance peak of $<\phi^{2}>_{H}$ is observed in the field-mass dependence. (If $\kappa>\kappa_{c}$, the polarization amplitude monotoneously decreases with increase of $m$.) Though the value of $\kappa_{c}$ remains uncertain within the analysis presented here, it is sure that dominant fields as quantum perturbations near the Schwarzschild horizon should be massless, while nearly extreme holes will have a quantum atmosphere dominated by fields with a resonance mass. The peak amplitude given by $8\pi^{2}r_{+}^{2}<\phi^{2}>_{max}\simeq0.0424$ at the nearly extreme Reissner-Nordstr\"{o}m horizon is not so smaller than the massless part given by $8\pi^{2}r_{+}^{2}<\phi^{2}>_{T}=1/6$ at the Schwarzschild horizon with the same area $4\pi r_{+}^{2}$. (If compared under the same black-hole mass $M$, the former becomes slightly larger than the latter evaluated by $8\pi^{2}M^{2}<\phi^{2}>_{T}=1/24$.) Considering a black hole evolving toward the zero-temperature state with a fixed radius $r_{+}$, we conclude that the mass $m$ of dominant fields generating vacuum polarization shifts from $mr_{+}\ll1$ to $mr_{+}\simeq0.38$ as the contribution of mass-induced excitation becomes important, without changing the polarization amplitude so much. Quantum back-reaction due to massive fields \cite{Taylor} will become very important for nearly extreme (low-temperature) black holes.

\acknowledgments

The authors wish to thank Y. Nambu for helpful discussions. This work was supported in part by the Grant in-aid for Scientific Research (C) of the Ministry of Education, Science, Sports and Culture of Japan (No.10640257).

\newpage


%
%

\end{document}